\documentclass[aoas,preprint]{imsart}

\RequirePackage{amsthm,amsmath,amsfonts,amssymb,mathtools}
\RequirePackage[authoryear]{natbib}
\RequirePackage[colorlinks,citecolor=blue,urlcolor=blue,backref=page]{hyperref}
\usepackage[ruled,vlined,linesnumbered]{algorithm2e}
\usepackage{bbm}

\RequirePackage{graphicx}
\usepackage{multirow}
\usepackage{float}
\usepackage{forest}
\usepackage{threeparttable}
\usepackage{caption}
\usepackage{subcaption}
\usepackage{eurosym}
\usepackage{etoolbox}
\usepackage{listings}
\usepackage{anyfontsize}
\usepackage{comment}
\usepackage{xcolor}
\AtBeginEnvironment{algorithm}{\linespread{0.8}\selectfont}
\pubyear{2024}
\volume{TBA}
\issue{TBA}
\firstpage{1}
\lastpage{1}
\startlocaldefs
\newcommand*\samethanks[1][\value{footnote}]{\footnotemark[#1]}
\theoremstyle{plain}

\theoremstyle{definition}

\theoremstyle{remark}

\endlocaldefs

\definecolor{codegreen}{rgb}{0,0.6,0}
\definecolor{codegray}{rgb}{0.5,0.5,0.5}
\definecolor{codeblue}{rgb}{0,0,1}
\definecolor{codepurple}{rgb}{0.58,0,0.82}
\definecolor{backcolour}{rgb}{0.9,0.9,0.8}

\lstdefinestyle{mystyle}{
  backgroundcolor=\color{backcolour}, commentstyle=\color{codegreen},
  keywordstyle=\color{black},
  numberstyle=\tiny\color{codegray},
  stringstyle=\color{codepurple},
  basicstyle=\ttfamily\footnotesize,
  breakatwhitespace=false,         
  breaklines=true,                 
  captionpos=b,                    
  keepspaces=true,                 
  numbers=left,                    
  numbersep=5pt,                  
  showspaces=false,                
  showstringspaces=false,
  showtabs=false,                  
  tabsize=2
}

\begin{document}

\begin{frontmatter}
\title{Nonparametric regression for cost-effectiveness analyses with observational data - a tutorial}
\runtitle{Nonparametric CEA}

\begin{aug}
\author[A]{\fnms{\textbf{Jonas}}~\snm{\textbf{Esser}}{\footnotesize\textsuperscript{\!,\!}}\ead[label=e1]{j.esser@vu.nl}\orcid{0009-0005-4964-5525}},
\author[B]{\fnms{Mateus}~\snm{Maia}\samethanks{\footnotesize\textsuperscript{\!,\!}}\ead[label=e2]{mateus.maiamarques@glasgow.ac.uk}\orcid{0000-0001-7056-386X}},
\author[A]{\fnms{\linebreak{}Judith E.}~\snm{Bosmans}\ead[label=e4]{j.e.bosmans@vu.nl}\orcid{0000-0002-1443-1026}}
\and
\author[A]{\fnms{Johanna Maria}~\snm{van Dongen}\ead[label=e5]{j.m.van.dongen@vu.nl}\orcid{0000-0002-1606-8742}}


\address[A]{Faculty of Science, Health Economics and Health Technology Assessment, Vrije Universiteit Amsterdam,\linebreak{}\printead[presep={}]{e1,e4,e5}}

\address[B]{School of Mathematics and Statistics, University of Glasgow,\printead[presep={}]{e2}}

\runauthor{Esser et al.}

\end{aug}
\begin{abstract}
Healthcare decision-making often requires selecting among multiple treatment options under budget constraints, particularly when one option is more effective but also more costly. Cost-effectiveness analysis (CEA) provides a framework for evaluating whether the health benefits of a treatment justify its additional costs. A key component of a CEA is the estimation of treatment effects on both health outcomes and costs, which becomes challenging when using observational data, due to potential confounding. While advanced causal inference methods exist for use in such circumstances, their adoption in CEAs remains limited, with many studies relying on overly simplistic methods such as linear regression or propensity score matching. 
In this paper, we address this gap by introducing cost-effectiveness researchers to modern nonparametric regression models, with a particular focus on Bayesian Additive Regression Trees (BART). We provide guidance on how to implement BART in CEAs, including code examples, and discuss its advantages in producing more robust and credible estimates from observational data. 
\end{abstract}

\end{frontmatter}

\textbf{Key points}
\begin{itemize}
    \item Estimating treatment effects in cost-effectiveness analyses (CEAs) requires careful adjustment for confounding, yet many studies still rely on linear regression and other simplistic methodology.
    \item This paper introduces Bayesian Additive Regression Trees (BART) as a flexible, nonparametric alternative for CEAs, offering practical guidance and code to help researchers generate more credible and robust estimates.
\end{itemize}

\section{Introduction}

A central challenge in healthcare decision-making is determining which interventions to implement or reimburse within the limits of constrained budgets. This often necessitates choosing between two or more competing treatment options for the same medical condition, where one treatment proves to be both more effective and more costly than the other. Cost-effectiveness analyses (CEAs) are designed to inform such choices by assessing whether the additional health gains of a treatment justify its additional costs.  A key component of a CEA is the estimation of the treatment effects on both patients’ health outcomes and associated costs using available data.

If cost-effectiveness data are obtained from a randomized trial, where treatment allocation does not depend on the patients' baseline characteristics, the statistical estimation of treatment effects is relatively simple. If the data are observational, however, and treatment allocation is hence related to the patients' baseline characteristics, the estimation procedure is much more complex. Despite the availability of several textbooks on the analysis of observational data \citep{imbens2015pate,aronow2019foundations,hernan2024causal}, appropriate methodology is not widely used in the field of CEA. That is, many researchers rely on simple linear regression or propensity score matching, preventing them from capturing complex, nonlinear relationships or interactions within the data. 
While there are some proposals for more sophisticated statistical methods in CEAs \citep{kreif2013regression,chen2024tutorial}, they still depend on restrictive parametric models, which may limit them from appropiately modeling complex data structures.

We see several potential reasons for the reliance on such relatively simple models. One is simply researchers' potential unawareness of useful alternatives, or of the advantages which these alternatives offer. Another is the added complexity of nonparametric models, with respect to both usage and interpretation.
In this paper, we aim to enlargen the toolkit available to cost-effectiveness researchers, by providing an accessible guideline on how to use state-of-the-art nonparametric regression methods for CEAs. Our focus will be on the \emph{Bayesian Additive Regression Trees} (BART) model, which has recently been extended for application in the cost-effectiveness setting. 

\section{Preliminaries}\label{sec:preliminaries}

We begin by explaining what researchers aim to learn from CEAs using observational data. Our presentation will be quite informal; a more rigorous version may be found in \citet{esser2024seemingly}. We shall not consider more general issues in the analysis of observational data, such as well-defined patient populations, treatment options, follow-up periods and so on. Such aspects are important, arguably more important than the statistical aspects, but they apply to CEAs in exactly the same way as they do to other epidemiological analyses. We thus refer the reader to canonical sources on these issues, such as \citet{hernan2016using}.

In a typical CEA, the researcher is faced with two treatment options, $t = 0$ and $t = 1$. Often the former is called the \emph{control}, and the latter the \emph{intervention}. The fundamental goal of a CEA is to estimate, given available data, the probability that the intervention is cost-effective, compared with the control. To make this possible, we consider two essential quantities:

\begin{itemize}
    \item $\Delta_c$, the \emph{average treatment effect on the costs}. Put simply, $\Delta_c$ tells us how much more we would need to pay per patient, on average, if every patient received the intervention instead of the control. It is typically expressed in the local currency which healthcare authorities are concerned with --- for instance, the Dutch National Health Care Institute uses EUR \citep{hakkaart2015costing}.
    \item $\Delta_q$, the \emph{average treatment effect on the patient's health\footnote{Instead of \emph{health} many authors use the term \emph{effect} here; see for example \citet{drummond2015methods}. We avoid this terminology, since we consider it confusing: $\Delta_q$ would then be the "effect on the effect".}}. This quantity tells us how much healthier the patients would be, on average, if every patient received the intervention instead of the control. The patients' health can be measured in many different ways. Some are disease-specific, others are intentionally non-specific, in order to make the health state of patients comparable across different diseases. Of the latter, the most commonly used is the \emph{Quality-adjusted life year} (QALY) \citep{drummond2015methods}.
\end{itemize}
As indicated above, we frequently encounter the situation that both treatment effects are positive, meaning that the intervention is both more effective and more expensive than the intervention. A natural question then arises: do the added health gains of the intervention justify the increase in costs? To answer this, we need to quantify how much we are willing to pay for an additional unit of health. This quantity is called the \emph{willingness-to-pay} (WTP) and is customarily assigned the symbol $\lambda$.
\\~\\
\textbf{Example 1.} Suppose that costs are measured in EUR, and $\Delta_c = 1000$; health is measured in QALYs, and $\Delta_q = 0.1$. Then, by implementing the intervention, we gain an average health benefit of $0.1$ QALYs per patient, while, having to pay an additional $1000$ EUR. We hence need to pay $10000$ EUR in exchange for one QALY gained (because $1000/0.1=10000$). The intervention is cost-effective if and only if $\lambda$ is greater than $10000$. In other words, the intervention is cost-effective if and only if $\lambda * 0.1 - 1000 > 0$.
\\~\\
The decision process in the preceding example can be formalized by working with the the \emph{incremental net benefit} (INB):
$$
\text{INB}_{\lambda} \coloneqq \lambda \Delta_q - \Delta_c.
$$

By putting a monetary value on health ($\lambda$) and checking if the benefits of the intervention outweigh its additional costs, we can decide wether the additional health we get from the intervention is worth the extra cost. More precisely: if the INB is zero or negative, we conclude that the intervention is not cost-effective and should hence not be implemented. Conversely, if the INB is positive, we consider the intervention to be cost-effective and worthy of implementation.

In most applications, the true values of $\Delta_c$ and $\Delta_q$ are unknown, and so is the true value of $\text{INB}_{\lambda}$ for any given $\lambda$. They thus must be estimated from available data, and there will inevitably be uncertainty about these estimates. In this paper, we adopt the Bayesian point of view when conceptualizing this uncertainty, which means that we treat all unknown quantities as random variables. This allows us to speak about the \emph{probability of cost-effectiveness}\footnote{In the frequentist or "classical" paradigm, this is not possible, and we would instead have to argue in terms of p-values. See \citet{lothgren2000definition}.}, which is given by $\Pr(\text{INB}_{\lambda} > 0)$, the probability that the INB is positive given a specific $\lambda$ value.

While there are many different methods to estimate $\Delta_c$ and $\Delta_q$, they are all ultimately based on empirical observations of patient's cost and health. These observations are often positively or negatively correlated, as illustrated by the following examples. 

\begin{itemize}
    \item \emph{Positive correlation:} \citet{stargardt2014measuring} found that, for patients suffering from acute myocardial infarction, higher use of healthcare resources is associated with higher survival rates. 
    \item \emph{Negative correlation:} \citep{dominguez2015cost} found that, in asthma care, increased disease severity (hence worse health) is associated with higher healthcare costs.
\end{itemize}

\begin{figure}
    \centering
    \includegraphics[width = 0.8\textwidth]{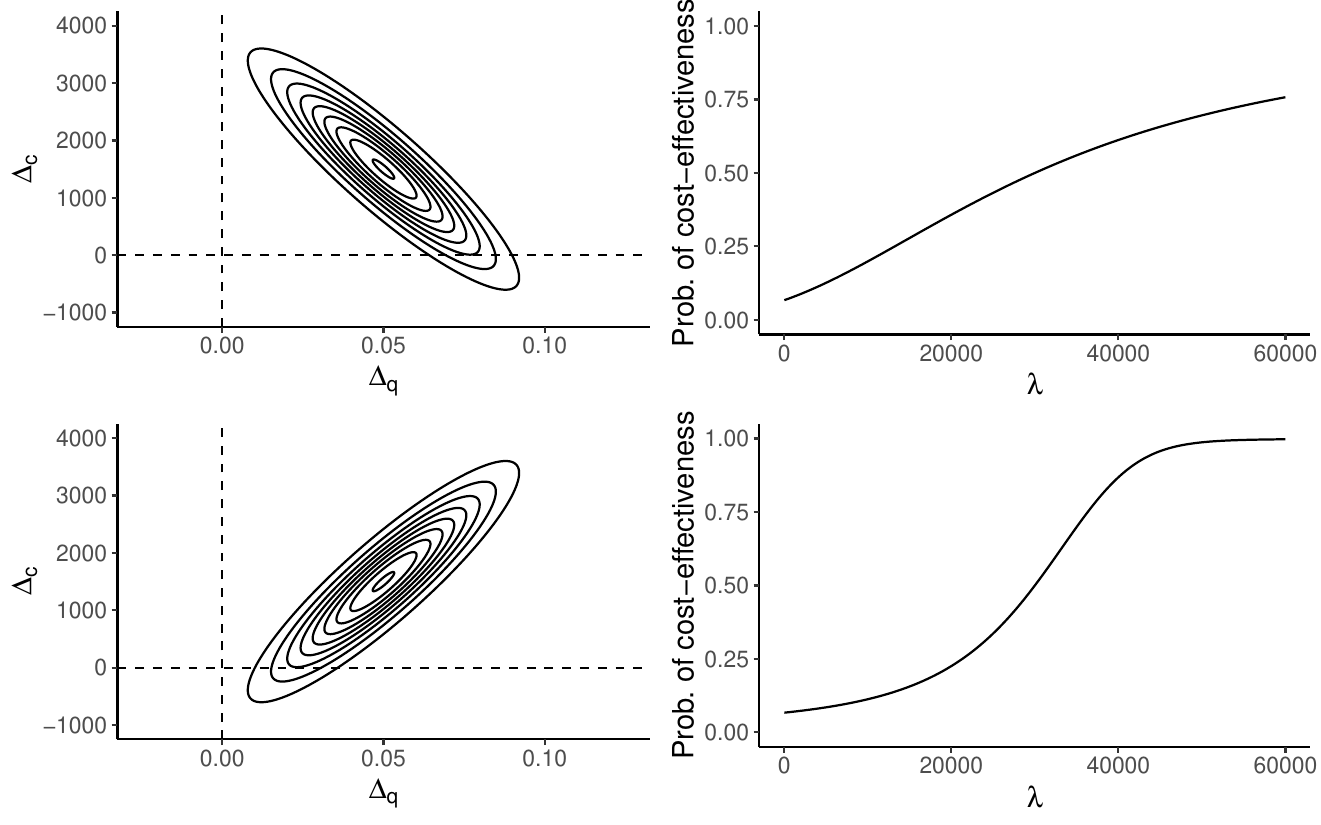}
    \caption{Correlation between $\Delta_c$ and $\Delta_q$ and its impact on the probability of cost-effectiveness.}
    \label{fig:cor_illustration}
\end{figure}

A proper statistical analysis should account for this correlation, and doing so will also induce correlation in the distribution of $\Delta_c$ and $\Delta_q$. The following example illustrates how this correlation can affect the results.
\\~\\
\textbf{Example 2.} We now consider a situation where $\Delta_c$ and $\Delta_q$ are estimated from available data, and we are hence uncertain about their values. This uncertainty is expressed through a probability distribution; an example of which is illustrated in Figure \ref{fig:cor_illustration}. The left column shows two bivariate density plots of $\Delta_q$ and $\Delta_c$, while the right column shows the probability of cost-effectiveness as a function of $\lambda$. These plots are called the \emph{cost-effectiveness plane} (CE-plane) and the \emph{cost-effectiveness acceptability curve} (CEAC), respectively. While the mean and variance of $\Delta_c$ and $\Delta_q$ are the same in both rows, the correlation between them differs: it is negative in the top row and positive in the bottom row. This difference in correlation leads to a noticeable change in the CEACs and, consequently, in how the results are interpreted. For example, at a willingness-to-pay of $50,000 \text{\euro}$ per QALY gained the probability of cost-effectiveness is approximately $0.7$ in the top row, but nearly $1$ in the bottom row.

\section{Observational studies and model specification}\label{sec:model_specification}

When cost-effectiveness data are obtained from a randomized trial, model specification is not critical. By design, the patients' baseline characteristics are unrelated to which treatment the patient receives, and we hence need not adjust for them; simply comparing mean costs and effects in the treatment groups will produce good estimates of the treatment effects. If a researcher insists on adjusting anyway, a simple linear model will perform well \citep{lin2013agnostic}.
Without randomization, however, there will generally be certain baseline characteristics, called \emph{confounders}, which influence both the treatment choice and the outcomes (i.e. cost and health) and hence need to be adjusted for in the analysis. This adjustment necessitates the use of a statistical model, as we need to specify how exactly the confounders influence the outcomes. Even with relevant confounders measured, a badly chosen model can lead to biased and inconsistent estimates of the treatment effects \citep{aronow2025nonparametric}, making model choice a critical matter when analysing observational data.

With that in mind, we will now critically examine some of the commonly used model choices in CEAs. A frequently used model is seemingly unrelated regression (SUR, \citet{zellner1962efficient, ben2023conducting}), which asserts that 
\begin{align}
\begin{pmatrix}  c_i \\ q_i \end{pmatrix} &= 
\begin{pmatrix}  \boldsymbol{\beta_1}^\top \mathbf{x_i}  \\ \boldsymbol{\beta_2}^\top \mathbf{x_i} \end{pmatrix} +
\begin{pmatrix}  \varepsilon_{i,1} \\ \varepsilon_{i,2} \end{pmatrix}, 
& \begin{pmatrix}  \varepsilon_{i,1} \\ \varepsilon_{i,2} \end{pmatrix} \sim \text{N}\Biggl(\begin{pmatrix}  0 \\ 0 \end{pmatrix}, \boldsymbol{\Sigma} \Biggl)
\label{eq:basic_SUR}
\end{align}

With SUR, we model the patients' costs $c_i$ and health $q_i$ as linear functions of a set of covariates $\mathbf{x_i}$, consisting the patients' baseline characteristics or some transformation thereof (so it may include, say, quadratic or interaction terms), plus a bivariate-normally distributed distributed error term. In a CEA, both the linearity and the normality assumption are questionable, and we shall now scrutinize each of them in turn.

The normality assumption is rarely realistic in the CEA setting, particularly for the cost outcome, which can only take non-negative values. However, violations of this assumption are not critical: in a landmark paper, \citet{white1980heteroskedasticity} showed that, as long as certain mild conditions are met, the estimates from linear regression models (such as SUR) remain unbiased and consistent\footnote{Consistency means, roughly speaking, that with a sufficiently large sample the estimate will be close to the true value.}, even when the distribution of the error term is misspecified. Nonetheless, the normality assumption can be weakened by through the use of bootstrapping, as is frequently done in CEA \citep{lothgren2000definition}.

Another way to avoid the normality assumption, which has been promoted by many authors but not applied much in practice, is the use of generalized linear models (GLMs). Those allow the specification of non-normal error terms—for example, by modeling costs using a lognormal or gamma distribution \citep{baio2018statistical,gabrio2019bayesian}. However, this approach has a significant drawback: the results can be very sensitive to the choice of the distribution. A valuable simulation study by \citet{briggs2005parametric} showed that, if a lognormal distribution is assumed for the costs, but the true distribution is gamma, the estimated mean costs are strongly biased. The sample mean on the other hand, which is essentially a linear
regression without covariates, performs well regardless of the true distribution. The same patterns will persist in the more observational setting, with adjustment for baseline covariates. The reason is that, for GLMs with non-normal error terms, there are no theoretical guarantees analogous to those in linear models — that is, if the error term in a GLM is not specified correctly, the resulting estimates may be inaccurate even with large sample sizes. In consequence, choosing the wrong distribution for the outcome can lead to both biased and inconsistent treatment effect estimates.

The more fundamental problem with the SUR model is the assumed linearity. Violations of this assumption can lead to severe bias in the estimated treatment effects \citep{dorie2019automated}; we illustrate this through a simple simulation experiment in Appendix \ref{app:simulation_experiment}. GLMs suffer from analogous problems. For example, a GLM for costs often assumes that the logarithm of the expected costs depends linearly on the covariates \citep{baio2018statistical}; an assumption just as restrictive as the linearity in the SUR model. More generally, both SUR and GLMs are examples of parametric models, which make strong assumptions about how the covariates are related to the outcomes. Violations of these assumptions lead to distorted estimates of treatment effects \citep{dorie2019automated}. 

In order to deal with the aforementioned problem, statisticians have developed regression techniques which do not make strong parametric assumptions about the relationship between the covariates and the outcomes. These so called \emph{nonparametric regression models}\footnote{The term "nonparametric" is somewhat of a misnomer, since such models do in fact have parameters. However, the number of these parameters (or more technically: the dimension of the parameter space) is not chosen in advance, unlike in, say, the linear model. Nonetheless, the terminology is widely used.} are often much more accurate than linear models or GLMs. Crucially, under similar regularity conditions, consistency guarantees analogous to those of linear regression,  also exist for nonparametric regression models, even when the normal error distribution is misspecified \citep{kleijn2006misspecification}.
These added benefits come at a cost, however, as nonparametric regression methods can be more challenging to use and interpret. In particular, unlike in the basic linear model, the treatment effects of interest are no longer represented by a single model parameter, making their interpretation les straightforward. We thus intend to make nonparametric regression accessible to applied cost-effectiveness researchers.

\section{BART for CEAs}

We will focus on a specific nonparametric regression model, namely \emph{Bayesian additive regression trees} (BART), which has been shown to perform well at estimating treatment effects  \citep{dorie2019automated}. 
We have recently extended this model to account for the correlation between costs and health outcomes found in CEA data; this extended version is called suBART \citep{esser2024seemingly}

As the name suggests, BART is a Bayesian statistical model and works by aggregating a set of regression trees. We shall now explain briefly what each of these terms mean. 

\subsection{Regression trees}
\begin{figure}[H]
\centering
\vspace{-0.5cm}
\begin{subfigure}[t]{0.44\textwidth}
\centering
        \begin{forest}
    for tree={
        grow=south, draw, minimum size=3ex, 
        inner sep=3pt, 
        s sep=7mm,
        l sep=6mm
    }
    [$x \le 0.3$,
        [$0.2$, edge label={node[midway, left, font=\footnotesize]{TRUE\:}}, circle,]
        [$x \le 0.6$, edge label={node[midway, right, font=\footnotesize]{\:FALSE}},
            [$x \le 0.5$, edge label={node[midway, left, font=\footnotesize]{}},
                [0.9, edge label={node[midway, left, font=\footnotesize]{}}, circle,]
                [0.4, edge label={node[midway, right, font=\footnotesize]{}}, circle,]
            ]
            [0.6, edge label={node[midway, right, font=\footnotesize]{}}, circle,]
        ]
    ]
\end{forest}
\end{subfigure}
\hspace*{\fill}
\begin{subfigure}[t]{0.54\textwidth}
\centering
     \includegraphics[width=0.8\columnwidth]{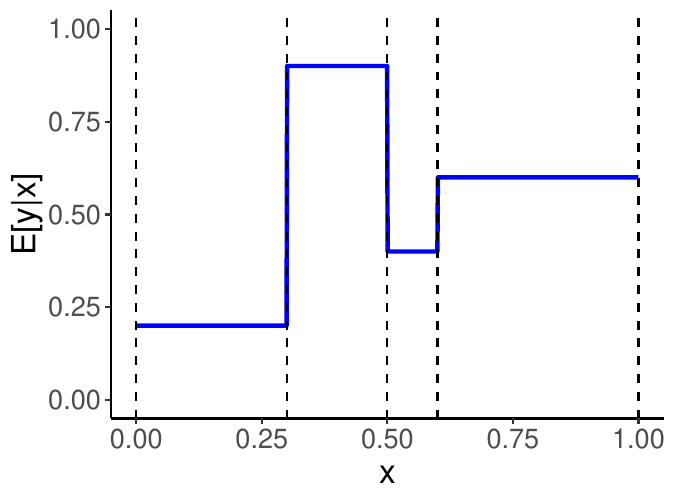}
\end{subfigure}
\caption{Regression tree (left) and resulting regression function (right).}
\label{fig:tree_example}
\end{figure}

While there regression model come in many forms, they all share the common goal of describing the relationship between an outcome variable and one or more covariates. A regression tree does this by dividing the covariate space (the set of all possible values of $\mathbf{x}_i$) into distinct "bins". Then each of these bins is assigned a single value, representing the predicted or expected outcome for all observations whose covariates fall within that bin. We illustrate this through a simple example.
\\~\\
\textbf{Example 3.} 
A representation of a simple regression tree is shown on the left of Figure \ref{fig:tree_example}. We may imagine this tree giving us the predicted QALY for a particular patient, based on their baseline health state (expressed as a number between 0 and 1). Given a specific value for this baseline utility --- say $x$ --- how do we find the predicted QALY? We start at the top of the tree, and then move down the levels sequentially. At each step, we follow either the left or the right direction, depending on whether $x$ is smaller or larger than the shown threshold value. Once we reach the end of a branch (where the tree no longer splits), we take the given value as the predicted QALY.
We can also use the tree to plot the expected QALY as a function of $x$. This is illustrated in the right panel of Figure \ref{fig:tree_example}. Where a linear model would have given us a line, the regression tree produces a so-called step function.
\\~\\
Regression trees are to some extent analogous to the decision tree models which health economists sometimes use in simulation experiments (see for example \citet{briggs2006decision}, chapter 2). However, among other differences, a regression tree is not a priori specified by the researcher, but rather "learned from the data".

\subsection{BART}

The basic idea of BART is to express the outcome as a sum of regression trees plus a normally distributed error term\footnote{As noted above, such a normally distributed error term does not pose issues even when the conditional distribution of the outcomes is not normal.}: 
\[y_i = \sum^m_{t=1} f_i\left(\mathbf{x}_i\right) + \varepsilon_i,\]
where each $f_i$ is a regression tree; each of which captures different nonlinear effects of the covariates on the outcome. By constraining the trees to be small, BART effectively controls overfitting, which in turn improves generalization. This framework allows for modelling complex nonlinear relationships between covariates and outcome without relying on strong parametric assumptions.
The total number of trees $m=200$ is frequently used, since this choice has been found to provide good performance while keeping the computational demands of the model fitting manageable. 
See \citet{chipman2010bart,tan2019bayesian} for detailed treatments of the original BART model, and particularly the process by which the trees are learned from the data.

\subsection{suBART}
The acronym suBART stands for \emph{seemingly unrelated BART}, a name deliberately reminiscent of the SUR model we described in Section \ref{sec:model_specification}. SUR consists of two linear models which are coupled through correlated error terms. suBART keeps the latter, but replaces the linear models with additive regression trees.
More precisely, the model is given by:
 \begin{align*}
\begin{pmatrix}  c_i \\ q_i \end{pmatrix} &= 
\begin{pmatrix}  \sum^m_{j = 1} f_j(\mathbf{x_i})   \\ \sum^m_{j = 1} g_j(\mathbf{x_i}) \end{pmatrix} +
\begin{pmatrix}  \varepsilon_{i,1} \\ \varepsilon_{i,2} \end{pmatrix}, 
& \begin{pmatrix}  \varepsilon_{i,1} \\ \varepsilon_{i,2} \end{pmatrix} \sim \text{N}\Biggl(\begin{pmatrix}  0 \\ 0 \end{pmatrix}, \boldsymbol{\Sigma} \Biggl),
\end{align*}
where each $f_i$ and $g_i$ is a regression tree.

\subsection{A note on Bayesian estimation and MCMC}
As the name suggests, BART is a Bayesian statistical method, which means that the model parameters --- and functions of these parameters, such as $\Delta_c$ --- are treated as random variables, rather than unknown constants. We learn about these parameters by examining their \emph{posterior distribution}, which quantifies how different values of the parameters are more or less likely, taking into account both prior beliefs and the evidence provided by the data.  

Bayesian estimation has a long tradition in CEA, with some authors arguing that it is the most natural framework to perform cost-effectiveness research under \citep{baio2018statistical}. The main reason is that cost-effectiveness researchers are concerned with the \emph{probability of cost-effectiveness} which is nothing other than $\Pr(\lambda \Delta_q - \Delta_c > 0)$. In other words, we want to make probabilistic statements about our parameters of interest, while explicitly acknowledging uncertainty about their values, and this -- strictly speaking --- is only possible under the Bayesian paradigm. 

In addition, there are pragmatic considerations that motivate our use of Bayesian estimation. It allows for the use of complex statistical models that are computationally difficult to estimate using alternative approaches. Morever, Bayesian estimation often performs well even when assessed by frequentist or other non-Bayesian criteria. BART is a prime example, as shown through the results of the American Causal Inference Conference (ACIC) competition \citep{dorie2019automated}.

In practice, Bayesian estimation works by taking a large number of random draws from the posterior distribution. We then use these draws to approximate our quantities of interest, such as the aforementioned probability of cost-effectiveness or confidence intervals for the treatment effects. This approximation will be quite accurate if the number of draws is sufficiently large. One popular procedure to obtain these draws is is called Markov chain Monte Carlo (MCMC), and the collection of draws is called an \emph{MCMC sample}. 
While superficially similar to bootstrapping, which is a commonly used tool in CEA, the underlying theory of MCMC is quite different. While bootstrap samples are resampled datasets used to approximate sampling variability in a frequentist framework, MCMC samples represent draws from the posterior distribution and quantify uncertainty in the parameters conditional on the observed data. In practice, however, the use of MCMC is quite similar to that of bootstrapping. For example, to construct a 90\% credible interval\footnote{The Bayesian analogue to the frequentist confidence interval.}, we can simply take the 5th and 95th percentiles of the MCMC sample.
Also analogously to bootstrapping, there are no strict guidelines for what the aforementioned "large number" of draws should be. The usual choice is a few thousand, and we will use $4000$ in the example below. 
For more information on MCMC, we refer to \citet{greenberg2012introduction}, Chapter 7.
Details on the specific MCMC method for the suBART model may be found in \citet{esser2024seemingly}.

\subsection{Propensity scores}
In the causal inference literature, one frequently encounters the so-called \emph{propensity score} (PS), which is the probability that a patient received the treatment, given their baseline characteristics. The PS summarizes all the confounding variables into a single number, and its inclusion as an additional covariate helps control for confounding more effectively \citep{hahn2020bayesian,dorie2019automated}. 
We therefore begin by estimating the PS for each patient in the data, using a BART model to predict treatment assignment based on baseline characteristics. This estimated PS is then included as an additional covariate when fitting the suBART model for the outcomes. This approach has been shown to perform well—and better than a suBART model without PS adjustment—in the context of observational CEAs \citep{esser2024seemingly}.

\section{Guided example}
In this section we guide the reader through the use of suBART. We use a simulated dataset, which is designed as an example and not meant to represent any specific real-world research endeavour. In the supplementary material\footnote{See \url{https://github.com/Jonas-Esser/nonparametric_CEA}}, we provide the full dataset, together with all of the following code.
The dataset contains the treatment indicator \textbf{t} (which indicates which treatment a patient received), the two outcome variables (costs and health --- represented by \textbf{c} and \textbf{q}, respectively) and the confounding variables \textbf{age}, \textbf{sex}, and \textbf{education}. For illustration, we display a subset of the data in Table~\ref{tab:data}.

\begin{table}[ht]
\caption{Subset of the simulated dataset used in the guided example. Variables include treatment ($\mathbf{t}$), outcomes ($\mathbf{c}$, $\mathbf{q}$), and covariates (\textbf{age},\textbf{sex} and \textbf{education}).}
\label{tab:data}
\centering
\begin{tabular}{rrrrrr}
\hline
\textbf{c} & \textbf{q} & \textbf{t} & \textbf{age} & \textbf{sex} & \textbf{education} \\
\hline
NA & 0.5865 & 0 & 56 & 0 & 3 \\
2882.69 & 0.9512 & 0 & 70 & 1 & 3 \\
2275.42 & 0.9149 & 0 & 71 & 1 & 3 \\
1964.08 & NA & 0 & 61 & 0 & 3 \\
2524.98 & 0.9133 & 0 & 75 & 1 & 1 \\
2683.89 & 0.6261 & 1 & 50 & 1 & 3 \\
NA & NA & 0 & 43 & 1 & 2 \\
1744.14 & 0.2123 & 1 & 21 & 1 & 3 \\
NA & 0.5122 & 1 & 28 & 0 & 3 \\
\ldots & \ldots & \ldots & \ldots & \ldots & \ldots \\
\hline
\end{tabular}
\end{table}

Some observations in the outcome variables are missing. The suBART method can handle these automatically, provided we assume that the data are \emph{missing at random} ---that is, the probability of a value being missing does not depend on the value itself, but may depend on other observed variables (such as confounders).

\subsection{Data formatting}
In order for the suBART code to work correctly, it is important that the data are formatted properly.
Unordered categorical variables with more than two levels (i.e., variables that represent categories without any natural order, such as blood type, region, or treatment center) should be declared as factors. In the dataset, this applies to the \texttt{education} variable. All other variables should be declared as numbers. 
Furthermore, all missing observations should be properly formatted as \texttt{NA} values, \emph{not} as text strings or numbers such as $-99$, which is occasionally observed in practice.

\begin{lstlisting}[language=R, caption={Loading the data, handling the unordered categorical variables and handle missing data in the outcomes, respectively.}]
rm(list=ls())
set.seed(42)

data_url <- "https://raw.githubusercontent.com/Jonas-Esser/nonparametric_CEA/refs/heads/main/data_example_missingness.csv"
data <- read.csv(data_url)

data$education <- as.factor(data$education)
data[data == -99] <- NA
\end{lstlisting}
\subsection{Data preparation}
We put all covariates into a data frame called \texttt{X}, including the treatment indicator but \emph{not} the outcomes \texttt{c} and \texttt{q}. The \texttt{X} object should contain only those variables which are supposed to enter into the regression model, and no others.
Likewise, we put the outcomes \texttt{c} and \texttt{q} into their own data frame \texttt{Y}.

\begin{lstlisting}[language=R, caption={Construction of covariate data frame \texttt{X} and outcome data frame \texttt{Y} by selecting relevant covariates and response variables}]
X <- data[,c("t", "age", "sex", "education")]
Y <- data[,c("c", "q")]
\end{lstlisting}
\subsection{Estimation of propensity scores}
We proceed to estimate the PS using a BART model, where treatment assignment is modeled as a function of the previously defined confounders. The number of trees and MCMC iterations was specified earlier. 
In line with standard practice, we discard the first 1000 posterior draws (usually referred to as burn-in samples) following standard practice in the MCMC literature. This means that although we set the total number of MCMC iterations (\texttt{n\_mcmc}) to $5000$ in the code, only the remaining $4000$ samples after burn-in are used for inference.

Once estimated, we append the PS to our covariate data frame, \texttt{X}. For clarity and to keep the original data intact, we store this extended data frame in a new object called \texttt{X\_ps}.

\begin{lstlisting}[language=R, caption= {Estimation of PS by fitting a BART model using the \texttt{subart} package. When the outcome is univariate, \texttt{subart} reduces to fitting a standard BART model.}]
library(subart)

ps_fit <- subart(x_train = X[,c("age", "sex", "education")],
                 y_mat = as.matrix(X$t),
                 n_tree = 100,
                 n_mcmc = 5000,
                 n_burn = 1000,
                 varimportance = TRUE)

X_ps <- X
X_ps$ps <- apply(pnorm(ps_fit$y_hat), 1, mean)
\end{lstlisting}

\subsection{Further data preparation}\label{sec:dataprep}
As mentioned above, in nonlinear models such as BART, treatment effects estimates are not represented by a single model parameter, unlike with the simple linear model without interaction terms. Instead, they  are derived from the model predictions as follows: for each patient, we estimate the expected outcomes under both treatment options and then calculate the average difference between these expected outcomes. The procedure is commonly known as \emph{g-computation} \citep{hernan2024causal}.

In algorithmic form, our strategy may be outlined as follows:

\begin{enumerate}
    \item Fit the suBART model.
    \item For each patient $i$, predict according to the model:
    \begin{itemize}
        \item $\hat{c}_{i}(0)$ and $\hat{q}_{i}(0)$, which are the expected costs and QALYs of the patient, \emph{had they received treatment $0$}, regardless of the treatment they actually received. 
        \item $\hat{c_i}(1)$ and $\hat{q_i}(1)$; defined analogously.
    \end{itemize}
    \item Obtain $\Delta_c$ and $\Delta_q$ by averaging the expected values over the sample:
    $$\Delta_c = \frac{1}{n} \sum^n_{i = 1} \hat{c_i}(1) - \hat{c_i}(0)$$
    $$\Delta_q = \frac{1}{n} \sum^n_{i = 1} \hat{q_i}(1) - \hat{q_i}(0)$$
    \item Using the obtained $\Delta_c$ and $\Delta_q$, compute the incremental net benefit for any willingness-to-pay $\lambda$ of interest, through
    $$\text{INB}_{\lambda} = \lambda \Delta_q - \Delta_c$$
    
\end{enumerate}
These steps will be performed for every MCMC draw.
For the sake of step $3$, we create another data frame, which contains each patient twice: once with the treatment indicator set to $0$, and once with it set to $1$. This data frame is called \texttt{X\_test} in the code. 
\begin{lstlisting}[language=R, caption= {Constructing the test dataset with treatment indicators to estimate potential outcomes under both treatment and control conditions.}]
X_test <- rbind(X_ps, X_ps)
X_test$t <- c(rep(0, nrow(X)), rep(1, nrow(X)))
\end{lstlisting}

\subsection{Fitting the suBART model}
We now fit the model, using the previously constructed data frames. 
\begin{lstlisting}[language=R, caption= Model fitting using suBART.] 
subart_fit <- subart(x_train = X_ps,
                        y_mat = as.matrix(Y),
                        x_test = X_test,
                        n_tree = 100,
                        n_mcmc = 5000,
                        n_burn = 1000,
                        varimportance = FALSE)
\end{lstlisting}
The resulting \texttt{suBART\_fit} object contains a variety of objects; we only need the entry \texttt{suBART\_fit\$y\_hat\_test}, which corresponds to the the predictions from Step \ref{sec:dataprep}. 

\subsection{Obtaining the results}
In this step, we apply the calculations from Section \ref{sec:dataprep} to all draws in the MCMC sample. We use the resulting draws of $\Delta_c$ and $\Delta_q$ to estimate $\Pr\left(\text{INB}_{\lambda} > 0\right)$ for a specified range of $\lambda$ values, which will allow us to compute and plot the cost-effectiveness acceptability curve (CEAC), as described by \citet{willan2006statistical}.
\begin{lstlisting}[language=R, caption= {Postprocessing of MCMC sample to compute average treatment effects, construct incremental $\text{INB}_{\lambda}$ at multiple willingness-to-pay thresholds, and calculate probabilities to build the CEAC}] 
n_post <- subart_fit$mcmc$n_mcmc - subart_fit$mcmc$n_burn
subart_fit$ATE <- matrix(NA, nrow = n_post, ncol = 2)

# Calculating Step 5.4.3;
for (i in 1:n_post){
  subart_fit$ATE[i,] <- c(
    mean(subart_fit$y_hat_test[X_test$t == 1,1,i]) - mean(subart_fit$y_hat_test[X_test$t == 0,1,i]),
    mean(subart_fit$y_hat_test[X_test$t == 1,2,i]) - mean(subart_fit$y_hat_test[X_test$t == 0,2,i])
    
  )
}

# Storing posterior MCMC sample 
MCMC_sample <- data.frame(
  Delta_c = subart_fit$ATE[,1],
  Delta_q = subart_fit$ATE[,2]
)
MCMC_sample$INB20 <- 20000 * MCMC_sample$Delta_q - MCMC_sample$Delta_c
MCMC_sample$INB50 <- 50000 * MCMC_sample$Delta_q - MCMC_sample$Delta_c

# Estimating probabilities for a specified range of lambda
lambda <- seq(0, 50000, length.out = 1000)
p <- rep(NA, length(lambda))
for (i in 1:length(lambda)){
  p[i] <- mean(lambda[i]*MCMC_sample$Delta_q - MCMC_sample$Delta_c > 0)
}
CEAC <- data.frame(lambda, p)
\end{lstlisting} \label{code:MCMC_sample}

\subsection{Plotting the results}
Finally, we plot the cost-effectiveness plane and CEAC, using the \texttt{ggplot2} package \citep{wickham2016ggplot}.
\begin{lstlisting}[language=R, caption= Plotting the cost-effectiveness plane and CEAC based on the posterior MCMC sample.]
library(ggplot2)

CEplane <- ggplot(MCMC_sample) + 
  geom_point(mapping = aes(Delta_q, Delta_c), size = 1, alpha = 0.3, show.legend = FALSE) +
  geom_point(aes(mean(Delta_q), mean(Delta_c)), color = "red", size = 2) +
  geom_vline(xintercept = 0, linetype = "dashed") +
  geom_hline(yintercept = 0, linetype = "dashed") +
  labs(x = expression(Delta[q]), y = expression(Delta[c])) +
  scale_x_continuous(limits = c(-0.05, 0.15)) +
  scale_y_continuous(limits = c(0, 1000)) +
  theme_classic() +
  theme(text = element_text(size = 14)) 

CEAC <- ggplot(data = CEAC) +
  geom_line(aes(x = lambda, y = p), linewidth = 0.8) +
  geom_vline(xintercept = 0, linetype = "dashed") +
  ylim(0,1) +
  labs(x = expression(lambda)) +
  labs(y = "Probability of cost-effectiveness") +
  theme_classic() +
  theme(text = element_text(size = 14), axis.line.y = element_blank(), legend.position = "right") +
  scale_x_continuous(label=paste0((0:5) * 10, "k")) +
  geom_vline(xintercept = 0) +
  geom_hline(yintercept = 0)

gridExtra::grid.arrange(CEplane, CEAC, nrow = 1)
\end{lstlisting}

\begin{figure}[htb]
    \centering
    \includegraphics[width = 1\textwidth]{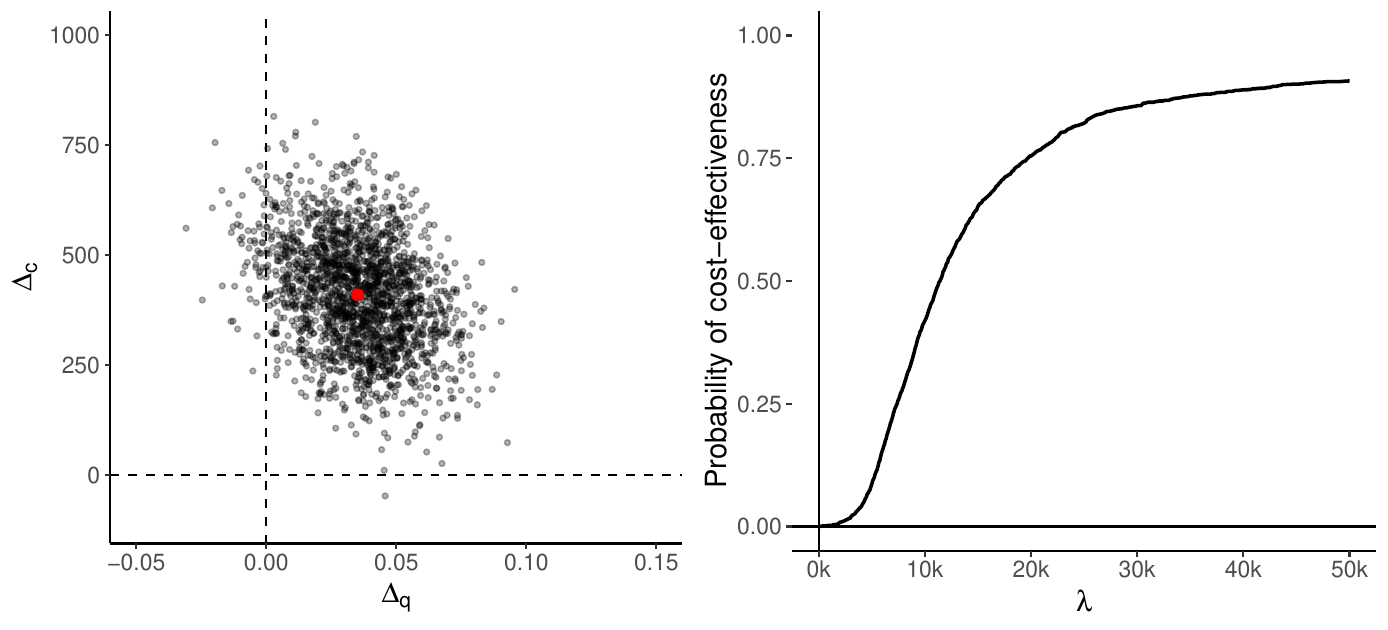}
    \caption{Cost-effectiveness plane (left) and CEAC (right).}
    \label{fig:CEplane}
\end{figure}

\section{Discussion} \label{seq:discussion}
In this tutorial, we demonstrated how to perform statistical estimation in CEAs with observational data. Our approach avoids strong parametric assumptions on the relationship between treatment and baseline variables with the expected costs and health state of a patient, which is very important when data is not obtained from a randomized trial. While we do not explicitly address the common issue of non-normal error terms, we argued that our approach is likely robust to violations of the normality assumption.

We used the suBART model in our presentation, both because of its strong empirical performance in this context and its methodological advantages.
We stress, however, that our primary goal was not to present this model, but rather to provide a general framework for cost-effectiveness researchers analyzing observational data. suBART is merely an example, and there are alternatives which can be used in a similar way, and which we also expect to perform well. In principle, any non-Bayesian method for estimating $\Delta_c$ and $\Delta_q$ can be used --- for example, the Causal Forest of \citet{wager2018estimation} or Double Machine Learning \citep{knaus2022double}. The MCMC sample referenced in Section \ref{code:MCMC_sample} would then be replaced by a bootstrap sample, and the rest of the analysis be carried out analogously. However, such an analysis would be no longer Bayesian, and hence the interpretation of the results changes: strictly speaking, we could no longer speak of the "probability of cost-effectiveness", and must instead argue in terms of frequentist p-values; see \citet{lothgren2000definition}. Additionally, such approaches may be infeasible in practice due to computational constraints.

We now discuss some limitations of our presented approach, relating to missing data, and outline possible solutions to address them.

\begin{itemize}
\item 
suBART can only deal with missing observations in the outcome variables, not in the covariates. The technical reason for this is that the model is formulated conditional on the covariates, and hence makes no assumptions about their distribution. In principle this is a positive aspect, as the analysis does then not rely on such assumptions, but it also implies that there is no way to impute missing covariates. 
Conceptually it seems most appealing to directly impute the missing covariates using the BART machinery, as proposed by \citet{xu2016sequential}. While applying this idea to suBART is conceptually straightforward, a computational implementation is currently not available. As a pragmatic alternative, we suggest combining suBART with multiple imputation in the same way as is commonly done when performing CEAs with trial data. The procedure can be sketched as follows: we create multiple complete datasets using standard imputation techniques and compute, for each of these datasets, the point estimates and covariance matrix of $\Delta_c$ and $\Delta_q$ (see Listing \ref{code:MCMC_sample}). These are then pooled using Rubin's rules \citep{rubin2018multiple}, and the pooled estimates may be used to construct the CEAC. In this context, we refer to \citet{harel2007multiple} for a general review of multiple imputation, and  to \citet{ben2023conducting} for specific guidance on multiple imputation in trial-based CEAs.

\item In our proposed framework, each patient's cost and health outcomes are either fully observed or entirely missing. This excludes the common situation of censored observations, where data are \emph{partially} missing. For example, patients may drop out of the study earlier than intended, and the probability of this dropout occurring may be related to the patients baseline characteristics as well as their health state and associated costs. \citet{chen2022net} proposed a framework to jointly address censoring and covariate adjustment in observational studies. However, their approach remains limited by its reliance on parametric regression models which must be specified correctly by the analyst. Their framework could in principle be combined with nonparametric methods for dealing with censoring --- for example, the BART-based method by \citet{sparapani2016nonparametric}. As in the previous bullet point, a computational implementation of such a method is currently lacking.  


\end{itemize}
Despite these limitations, the presented framework is sufficient for use in many CEAs with observational data and produces more credible results than commonly used parametric models. We thus hope that readers will find it useful in their research. 

\section{Declarations}

\subsection{Funding}
No funding was received for the completion of this paper.

\subsection{Conflicts of interest}
All authors have no conflicts of interest to declare.

\subsection{Availability of data and materials}
All data and code used are available under \url{https://github.com/Jonas-Esser/nonparametric_CEA}.

\subsection{Ethics approval}
Not applicable.

\subsection{Author contributions}
The initial idea for the paper was proposed by Jonas Esser; the scope and design were then worked out in cooperation with Johanna van Dongen and Judith Bosmans. The text was written primarily by Jonas Esser, while incorporating suggestions from all other authors. The \texttt{R} code, figures and computer experiments were created by Jonas Esser and Mateus Maia.

\interlinepenalty=1000
\bibliographystyle{ba}
\bibliography{sample}

\begin{thebibliography}{36}
\newcommand{\enquote}[1]{``#1''}
\expandafter\ifx\csname natexlab\endcsname\relax\def\natexlab#1{#1}\fi
\expandafter\ifx\csname url\endcsname\relax
  \def\url#1{{\tt #1}}\fi
\expandafter\ifx\csname urlprefix\endcsname\relax\def\urlprefix{URL }\fi
\ifx\endbibitem\undefined \let\endbibitem\relax\fi

\bibitem[{Aronow et~al.(2025)Aronow, Robins, Saarinen, S{\"a}vje, and Sekhon}]{aronow2025nonparametric}
Aronow, P., Robins, J.~M., Saarinen, T., S{\"a}vje, F., and Sekhon, J. (2025).
\newblock \enquote{Nonparametric identification is not enough, but randomized controlled trials are.}
\newblock {\em Observational Studies\/}, 11(1): 3--16.
\endbibitem

\bibitem[{Aronow and Miller(2019)}]{aronow2019foundations}
Aronow, P.~M. and Miller, B.~T. (2019).
\newblock {\em Foundations of agnostic statistics\/}.
\newblock Cambridge University Press.
\endbibitem

\bibitem[{Baio(2018)}]{baio2018statistical}
Baio, G. (2018).
\newblock \enquote{Statistical modeling for health economic evaluations.}
\newblock {\em Annual review of statistics and its application\/}, 5(1): 289--309.
\endbibitem

\bibitem[{Ben et~al.(2023)Ben, van Dongen, El~Alili, Esser, Broul{\'\i}kov{\'a}, and Bosmans}]{ben2023conducting}
Ben, {\^A}.~J., van Dongen, J.~M., El~Alili, M., Esser, J.~L., Broul{\'\i}kov{\'a}, H.~M., and Bosmans, J.~E. (2023).
\newblock \enquote{Conducting trial-based economic evaluations using R: a tutorial.}
\newblock {\em Pharmacoeconomics\/}, 41(11): 1403--1413.
\endbibitem

\bibitem[{Briggs et~al.(2005)Briggs, Nixon, Dixon, and Thompson}]{briggs2005parametric}
Briggs, A., Nixon, R., Dixon, S., and Thompson, S. (2005).
\newblock \enquote{Parametric modelling of cost data: some simulation evidence.}
\newblock {\em Health economics\/}, 14(4): 421--428.
\endbibitem

\bibitem[{Briggs et~al.(2006)Briggs, Sculpher, and Claxton}]{briggs2006decision}
Briggs, A., Sculpher, M., and Claxton, K. (2006).
\newblock {\em Decision modelling for health economic evaluation\/}.
\newblock Oup Oxford.
\endbibitem

\bibitem[{Chen et~al.(2024)Chen, Bang, and Hoch}]{chen2024tutorial}
Chen, S., Bang, H., and Hoch, J.~S. (2024).
\newblock \enquote{A Tutorial on Net Benefit Regression for Real-World Cost-Effectiveness Analysis Using Censored Data from Randomized or Observational Studies.}
\newblock {\em Medical Decision Making\/}, 44(3): 239--251.
\endbibitem

\bibitem[{Chen and Hoch(2022)}]{chen2022net}
Chen, S. and Hoch, J.~S. (2022).
\newblock \enquote{Net-benefit regression with censored cost-effectiveness data from randomized or observational studies.}
\newblock {\em Statistics in medicine\/}, 41(20): 3958--3974.
\endbibitem

\bibitem[{Chipman et~al.(2010)Chipman, George, and McCulloch}]{chipman2010bart}
Chipman, H.~A., George, E.~I., and McCulloch, R.~E. (2010).
\newblock \enquote{BART: {B}ayesian additive regression trees.}
\newblock {\em The Annals of Applied Statistics\/}, 4(1): 266--298.
\endbibitem

\bibitem[{Dom{\'\i}nguez-Ortega et~al.(2015)Dom{\'\i}nguez-Ortega, Phillips-Angl{\'e}s, Barranco, and Quirce}]{dominguez2015cost}
Dom{\'\i}nguez-Ortega, J., Phillips-Angl{\'e}s, E., Barranco, P., and Quirce, S. (2015).
\newblock \enquote{Cost-effectiveness of asthma therapy: a comprehensive review.}
\newblock {\em Journal of Asthma\/}, 52(6): 529--537.
\endbibitem

\bibitem[{Dorie et~al.(2019)Dorie, Hill, Shalit, Scott, and Cervone}]{dorie2019automated}
Dorie, V., Hill, J.~L., Shalit, U., Scott, M., and Cervone, D. (2019).
\newblock \enquote{Automated versus do-it-yourself methods for causal inference: lessons learned from a data analysis competition.}
\newblock {\em Statistical Science\/}, 34(1): 43--68.
\endbibitem

\bibitem[{Drummond et~al.(2015)Drummond, Sculpher, Claxton, Stoddart, and Torrance}]{drummond2015methods}
Drummond, M.~F., Sculpher, M.~J., Claxton, K., Stoddart, G.~L., and Torrance, G.~W. (2015).
\newblock {\em Methods for the Economic Evaluation of Health Care Programmes\/}.
\newblock Oxford, UK: Oxford University Press, 4th edition.
\endbibitem

\bibitem[{Esser et~al.(2024)Esser, Maia, Parnell, Bosmans, van Dongen, Klausch, and Murphy}]{esser2024seemingly}
Esser, J., Maia, M., Parnell, A.~C., Bosmans, J., van Dongen, H., Klausch, T., and Murphy, K. (2024).
\newblock \enquote{Seemingly unrelated Bayesian additive regression trees for cost-effectiveness analyses in healthcare.}
\newblock {\em arXiv preprint arXiv:2404.02228\/}.
\endbibitem

\bibitem[{Gabrio et~al.(2019)Gabrio, Baio, and Manca}]{gabrio2019bayesian}
Gabrio, A., Baio, G., and Manca, A. (2019).
\newblock \enquote{Bayesian statistical economic evaluation methods for health technology assessment.}
\newblock In Hamilton, J. (ed.), {\em Economic Theory and Mathematical Models\/}. Oxford, UK: Oxford Research Encyclopedia of Economics and Finance.
\endbibitem

\bibitem[{Greenberg(2012)}]{greenberg2012introduction}
Greenberg, E. (2012).
\newblock {\em Introduction to {B}ayesian Econometrics\/}.
\newblock Cambridge, UK: Cambridge University Press.
\endbibitem

\bibitem[{Hahn et~al.(2020)Hahn, Murray, and Carvalho}]{hahn2020bayesian}
Hahn, P.~R., Murray, J.~S., and Carvalho, C.~M. (2020).
\newblock \enquote{Bayesian regression tree models for causal inference: regularization, confounding, and heterogeneous effects (with discussion).}
\newblock {\em Bayesian Analysis\/}, 15(3): 965--1056.
\endbibitem

\bibitem[{Hakkaart-van Roijen et~al.(2015)Hakkaart-van Roijen, Van~der Linden, Bouwmans, Kanters, Tan et~al.}]{hakkaart2015costing}
Hakkaart-van Roijen, L., Van~der Linden, N., Bouwmans, C., Kanters, T., Tan, S., et~al. (2015).
\newblock \enquote{Costing manual: Methodology of costing research and reference prices for economic evaluations in healthcare.}
\newblock {\em Diemen: Zorginstituut Nederland\/}.
\endbibitem

\bibitem[{Harel and Zhou(2007)}]{harel2007multiple}
Harel, O. and Zhou, X.-H. (2007).
\newblock \enquote{Multiple imputation: review of theory, implementation and software.}
\newblock {\em Statistics in Medicine\/}, 26(16): 3057--3077.
\endbibitem

\bibitem[{Hern{\'a}n and Robins(2016)}]{hernan2016using}
Hern{\'a}n, M.~A. and Robins, J.~M. (2016).
\newblock \enquote{Using big data to emulate a target trial when a randomized trial is not available.}
\newblock {\em American journal of epidemiology\/}, 183(8): 758--764.
\endbibitem

\bibitem[{Hernán and Robins(2024)}]{hernan2024causal}
Hernán, M.~A. and Robins, J.~M. (2024).
\newblock {\em Causal Inference: What If\/}.
\newblock Boca Raton, FL, U.S.A.: Chapman \& Hall/CRC Press.
\endbibitem

\bibitem[{Imbens and Rubin(2015)}]{imbens2015pate}
Imbens, G.~W. and Rubin, D.~B. (2015).
\newblock {\em Causal Inference for Statistics, Social, and Biomedical Sciences: An Introduction\/}.
\newblock New York, NY, U.S.A.: Cambridge University Press.
\endbibitem

\bibitem[{Kleijn and van~der Vaart(2006)}]{kleijn2006misspecification}
Kleijn, B. and van~der Vaart, A. (2006).
\newblock \enquote{Misspecification in infinite-dimensional Bayesian statistics.}
\newblock {\em The Annals of Statistics\/}, 34(2).
\endbibitem

\bibitem[{Knaus(2022)}]{knaus2022double}
Knaus, M.~C. (2022).
\newblock \enquote{Double machine learning-based programme evaluation under unconfoundedness.}
\newblock {\em The Econometrics Journal\/}, 25(3): 602--627.
\endbibitem

\bibitem[{Kreif et~al.(2013)Kreif, Grieve, Radice, and Sekhon}]{kreif2013regression}
Kreif, N., Grieve, R., Radice, R., and Sekhon, J.~S. (2013).
\newblock \enquote{Regression-adjusted matching and double-robust methods for estimating average treatment effects in health economic evaluation.}
\newblock {\em Health Services and Outcomes Research Methodology\/}, 13: 174--202.
\endbibitem

\bibitem[{Lin(2013)}]{lin2013agnostic}
Lin, W. (2013).
\newblock \enquote{Agnostic notes on regression adjustments to experimental data: Reexamining Freedman’s critique.}
\newblock {\em The Annals of Applied Statistics\/}, 7(1).
\endbibitem

\bibitem[{L{\"o}thgren and Zethraeus(2000)}]{lothgren2000definition}
L{\"o}thgren, M. and Zethraeus, N. (2000).
\newblock \enquote{Definition, interpretation and calculation of cost-effectiveness acceptability curves.}
\newblock {\em Health Economics\/}, 9(7): 623--630.
\endbibitem

\bibitem[{Rubin(2018)}]{rubin2018multiple}
Rubin, D.~B. (2018).
\newblock \enquote{Multiple imputation.}
\newblock In {\em Flexible imputation of missing data, second edition\/}, 29--62. Chapman and Hall/CRC.
\endbibitem

\bibitem[{Sparapani et~al.(2016)Sparapani, Logan, McCulloch, and Laud}]{sparapani2016nonparametric}
Sparapani, R.~A., Logan, B.~R., McCulloch, R.~E., and Laud, P.~W. (2016).
\newblock \enquote{Nonparametric survival analysis using Bayesian additive regression trees (BART).}
\newblock {\em Statistics in medicine\/}, 35(16): 2741--2753.
\endbibitem

\bibitem[{Stargardt et~al.(2014)Stargardt, Schrey{\"o}gg, and Kondofersky}]{stargardt2014measuring}
Stargardt, T., Schrey{\"o}gg, J., and Kondofersky, I. (2014).
\newblock \enquote{Measuring the relationship between costs and outcomes: the example of acute myocardial infarction in German hospitals.}
\newblock {\em Health Economics\/}, 23(6): 653--669.
\endbibitem

\bibitem[{Tan and Roy(2019)}]{tan2019bayesian}
Tan, Y.~V. and Roy, J. (2019).
\newblock \enquote{Bayesian additive regression trees and the General {BART} model.}
\newblock {\em Statistics in Medicine\/}, 38(25): 5048--5069.
\endbibitem

\bibitem[{Wager and Athey(2018)}]{wager2018estimation}
Wager, S. and Athey, S. (2018).
\newblock \enquote{Estimation and inference of heterogeneous treatment effects using random forests.}
\newblock {\em Journal of the American Statistical Association\/}, 113(523): 1228--1242.
\endbibitem

\bibitem[{White(1980)}]{white1980heteroskedasticity}
White, H. (1980).
\newblock \enquote{A heteroskedasticity-consistent covariance matrix estimator and a direct test for heteroskedasticity.}
\newblock {\em Econometrica: journal of the Econometric Society\/}, 817--838.
\endbibitem

\bibitem[{Wickham(2016)}]{wickham2016ggplot}
Wickham, H. (2016).
\newblock {\em ggplot2: Elegant Graphics for Data Analysis\/}.
\newblock Springer-Verlag New York.
\newline\urlprefix\url{https://ggplot2.tidyverse.org}
\endbibitem

\bibitem[{Willan and Briggs(2006)}]{willan2006statistical}
Willan, A.~R. and Briggs, A.~H. (2006).
\newblock {\em Statistical Analysis of Cost-Effectiveness Data\/}.
\newblock Statistics in Practice. John Wiley \& Sons.
\endbibitem

\bibitem[{Xu et~al.(2016)Xu, Daniels, and Winterstein}]{xu2016sequential}
Xu, D., Daniels, M.~J., and Winterstein, A.~G. (2016).
\newblock \enquote{Sequential BART for imputation of missing covariates.}
\newblock {\em Biostatistics\/}, 17(3): 589--602.
\endbibitem

\bibitem[{Zellner(1962)}]{zellner1962efficient}
Zellner, A. (1962).
\newblock \enquote{An efficient method of estimating seemingly unrelated regressions and tests for aggregation bias.}
\newblock {\em Journal of the American Statistical Association\/}, 57(298): 348--368.
\endbibitem

\end{thebibliography}



\clearpage
\setcounter{subsection}{0}
\renewcommand\thesubsection{\Alph{subsection}}
\renewcommand\thefigure{\thesubsection.\arabic{figure}}
\renewcommand\thetable{\thesubsection.\arabic{table}}

\section*{Appendix}\label{sec:appendix}

\subsection{Simulation experiment}\label{app:simulation_experiment}
We demonstrate here, through a simple simulation experiment, the consequences of model misspecification when estimating treatment effects. Such experiments are useful because we know the true data-generating process (hence also the true treatment effects) and can hence investigate how a statistical method performs at recovering the truth. The \texttt{R} code to replicate the experiment is given at the end.

The goal of our hypothetical analysis is estimating the effect of a treatment $t_i$ on the healthcare costs $c_i$, and there is a single continuous baseline covariate $x_i$. Our true model is given by 
\begin{equation*}\label{eq:true_model}
c_i = t_i + 3\sin(x_i) x_i^2 + \epsilon_i.
\end{equation*}
In line with common practice, our statistical model of choice is a simple linear regression:
\begin{equation}\label{eq:assumed_model}
c_i = \beta_0 + \beta_1 t_i + \beta_2 x_i + \epsilon_i
\end{equation}
where $\epsilon_i \sim \mathbf{N}(0,\sigma^2).$ Under the assumption of correct model specification, we have $\beta_1 = \Delta_c$.

However, our model in Equation~\eqref{eq:assumed_model} is evidently misspecified. This raises the question of what happens to our estimates when we fit model ~\eqref{eq:assumed_model} despite its  misspecification. Specifically, we ask whether the estimates can still yield meaningful information about the underlying treatment effect.

As mentioned in Section \ref{sec:model_specification}, the answer depends on whether the treatment assignment depends on the baseline covariate $x_i$. We first assign to each patient a probability of $0.5$ of receiving treatment $t_i = 1$, independently of $x_i$. The simulation setup hence mimics a randomized trial. 

\begin{figure}[htb]
    \centering
    \includegraphics[width = 0.8\textwidth]{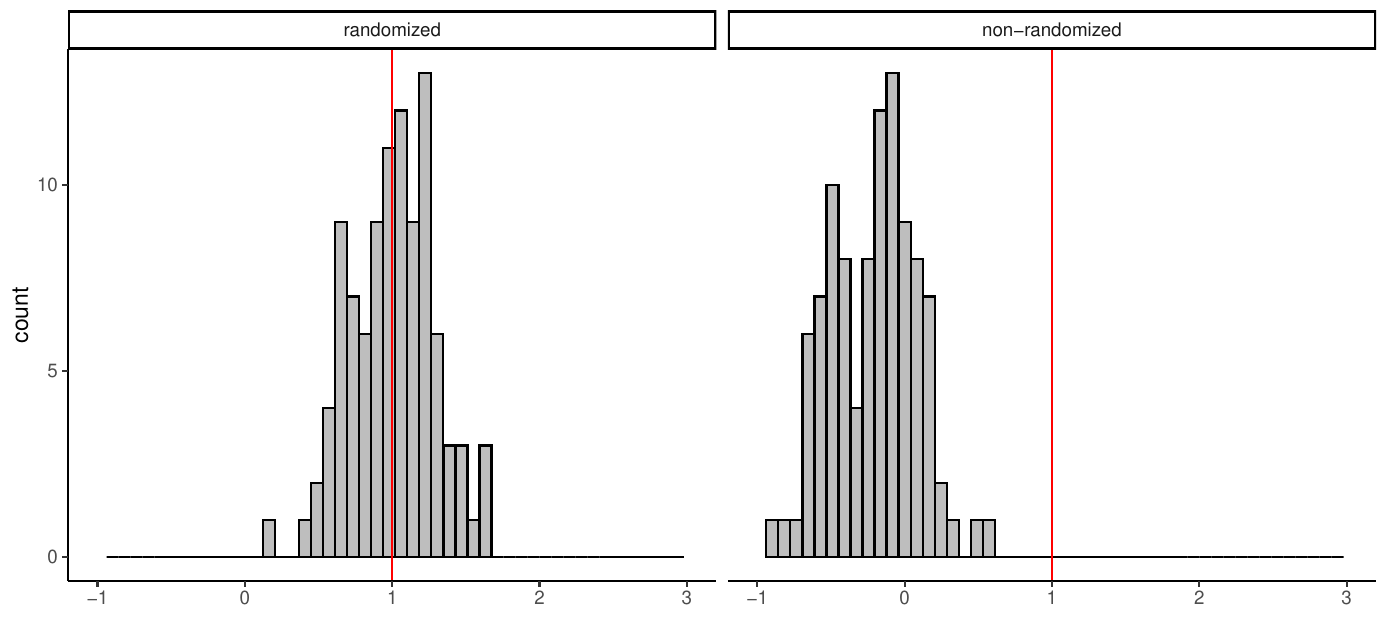}
    \caption{Estimates of $\Delta_c$, obtained through linear regression.}
    \label{fig:hist_linear}
\end{figure}

We now draw $100$ random datasets from the specified simulation and analyse each dataset using the linear regression model. The left panel in Figure~\ref{fig:hist_linear} shows a histogram of the estimated treatment effects. Despite the misspecification, the estimates concentrate around the true value $1$.

We now modify the simulation such that $t_i$ is no longer independent of $x_i$, thereby mimicking an observational study, where the treatment assignment depends on the patients' baseline characteristics. 
\begin{equation}
    \text{Pr}(t_i = 1)= \frac{1}{1 + e^{-3x_i}}
    \label{eq:nonlinear_ps}
\end{equation}
The probability in Equation \eqref{eq:nonlinear_ps} is called the \emph{propensity score} (PS).

The results are shown in the right panel of Figure \ref{fig:hist_linear}. Unlike before, we now observe a clear bias in the estimates. 

While our example was of course rather simplistic, the same patterns can be observed in more complex and realistic simulation experiments. A case in point is the seminal work by \citet{dorie2019automated}, where a much more elaborate and systematic simulation experiment may be found. The authors simulated data from a variety of hypothetical observational studies, using nonlinear data-generating processes. They then tasked other teams of researchers, who were unaware of the true data-generating processes, to try to estimate the average treatment effect through a statistical method of their choice. It turned out that \emph{all} variations of linear regression models performed poorly, as did other kinds of parametric models, whereas some nonparametric regression methods performed well. 

\begin{figure}
    \centering
    \includegraphics[width = 0.8\textwidth]{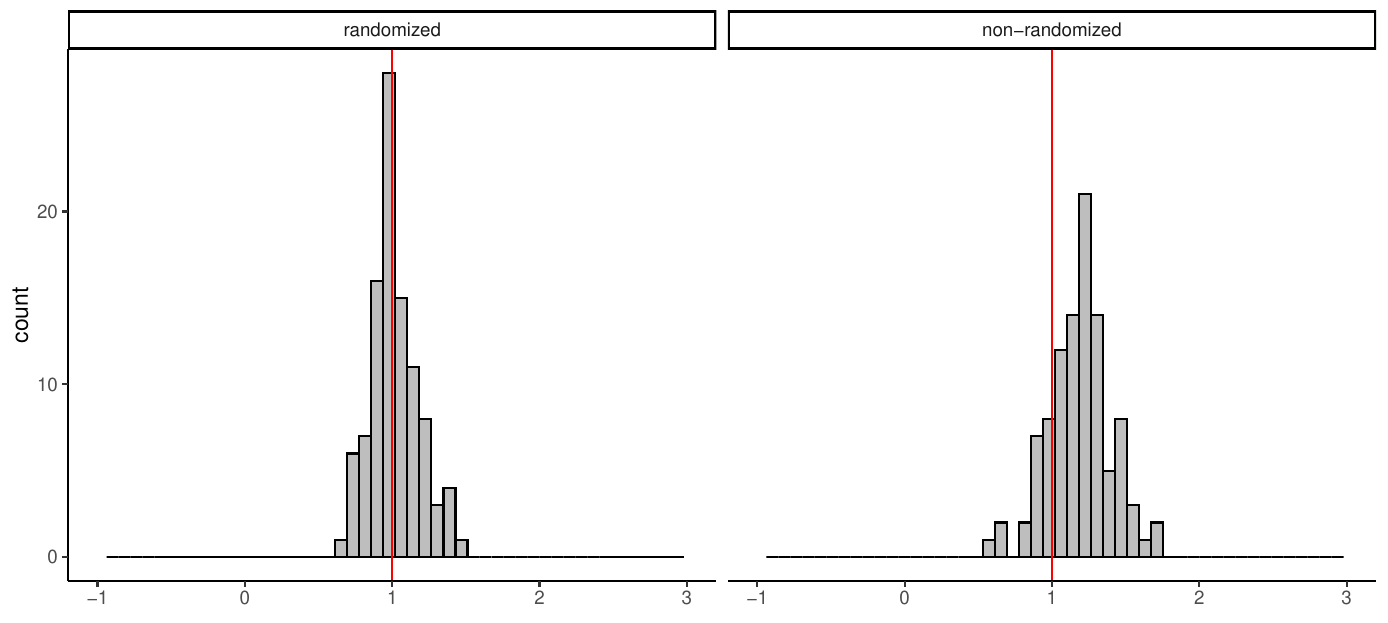}
    \caption{Estimates of $\Delta_c$, obtained through BART.}
    \label{fig:hist_BART}
\end{figure}

To illustrate the utility of nonparametric regression, we rerun the simulation experiment from the previous section, using BART instead of a linear regression. The results are presented in \ref{fig:hist_BART}. For each of the two data-generating processes, the inferred treatment effects concentrate around the true value, which shows that BART can correctly estimate the treatment effects even when the true relationship is nonlinear.

\begin{lstlisting}[language=R, caption= {Code for the simulation experiment}]
rm(list=ls())
library(subart)
library(ggplot2)

N_sim <- 100
N <- 200
X <- rnorm(N)
ATE_linear_sim1 <- rep(NA, N_sim)
ATE_linear_sim2 <- rep(NA, N_sim)
ATE_BART_sim1 <- rep(NA, N_sim)
ATE_BART_sim2 <- rep(NA, N_sim)
EY_0 <- 3 * sin(X) * X^2
EY_1 <- EY_0 + 1
ATE_true <- mean(EY_1 - EY_0)

test <- matrix(c(rep(X, 2), rep(0, N), rep(1, N)), ncol = 2)
test <- as.data.frame(test)

# sim 1: random assignment
for (i in 1:N_sim){
     A <- rbinom(N, 1, 0.5)
     Y <- (1 - A)*EY_0 + A*EY_1 + rnorm(N)
     ATE_linear_sim1[i] <- lm(Y ~ A + X)$coefficients[2]
     train <- data.frame(X,A)
     BART_mod <- subart(x_train = train,
                        y_mat = as.matrix(Y),
                        x_test = test,
                        n_tree = 100,
                        n_mcmc = 1100,
                        n_burn = 100,
                        varimportance = FALSE)
     ATE_BART_sim1[i] <- mean(BART_mod$y_hat_test_mean[(N+1):(2*N),1] - BART_mod$y_hat_test_mean[1:N,1])
}

# sim 2: nonrandom assigment
for (i in 1:N_sim){
     A <- rbinom(N, 1, 1/(1 + exp(-3*X)))
     Y <- (1 - A)*EY_0 + A*EY_1 + rnorm(N)
     ATE_linear_sim2[i] <- lm(Y ~ A + X)$coefficients[2]
     train <- data.frame(X,A)
     BART_mod <- subart(x_train = train,
                        y_mat = as.matrix(Y),
                        x_test = test,
                        n_tree = 100,
                        n_mcmc = 1100,
                        n_burn = 100,
                        varimportance = FALSE)
     ATE_BART_sim2[i] <- mean(BART_mod$y_hat_test_mean[(N+1):(2*N),1] - BART_mod$y_hat_test_mean[1:N,1])
}

data_sim <- data.frame(sim = c(rep("randomized",N_sim),rep("non-randomized",N_sim)),
                       ATE_linear = c(ATE_linear_sim1,ATE_linear_sim2),
                       ATE_BART = c(ATE_BART_sim1,ATE_BART_sim2)
)
data_sim <- data.frame(sim = c(rep("randomized",N_sim),rep("non-randomized",N_sim)),
                       ATE_linear = c(ATE_linear_sim1,ATE_linear_sim2),
                       ATE_BART = c(ATE_BART_sim1,ATE_BART_sim2)
)
data_sim$sim <- factor(data_sim$sim, levels = c("randomized", "non-randomized"))

ggplot(data_sim) +
     geom_histogram(aes(x = ATE_linear), color = "black", fill = "grey") +
     scale_x_continuous(limits = c(-1,3)) +
     geom_vline(xintercept = 1, color = "red") +
     facet_wrap(. ~ sim) +
     theme_classic() +
     theme(axis.title.x = element_blank())

ggplot(data_sim) +
     geom_histogram(aes(x = ATE_BART), color = "black", fill = "grey") +
     scale_x_continuous(limits = c(-1,3)) +
     geom_vline(xintercept = 1, color = "red") +
     facet_wrap(. ~ sim) +
     theme_classic() +
     theme(axis.title.x =  element_blank())


\end{lstlisting}

\end{document}